\documentclass[twocolumn,showpacs,superscriptaddress,preprintnumbers,amsmath,amssymb,prc]{revtex4-2}
\usepackage{graphicx}
\usepackage{dcolumn}

\usepackage{bm}

\usepackage{CJKutf8}
\usepackage{upgreek} 
\usepackage{subfigure}
\usepackage{makecell}
\usepackage{xcolor} 

\usepackage[colorlinks,linkcolor=blue,urlcolor=blue,citecolor=blue]{hyperref}
\usepackage{tikz,xcolor,hyperref}

\definecolor{lime}{HTML}{A6CE39}
\DeclareRobustCommand{\orcidicon}{
	\begin{tikzpicture}
	\draw[lime, fill=lime] (0,0) 
	circle [radius=0.16] 
	node[white] {{\fontfamily{qag}\selectfont \tiny ID}};
	\draw[white, fill=white] (-0.0625,0.095) 
	circle [radius=0.007];
	\end{tikzpicture}
	\hspace{-2mm}
}
\foreach \x in {A, ..., Z}{%
	\expandafter\xdef\csname orcid\x\endcsname{\noexpand\href{https://orcid.org/\csname orcidauthor\x\endcsname}{\noexpand\orcidicon}}
}
\foreach \x in {A, ..., Z}{%
	\expandafter\xdef\csname orcid\x\endcsname{\noexpand\href{https://orcid.org/\csname orcidauthor\x\endcsname}{\noexpand\orcidicon}}
}

\begin{document}

\title{New measurement of $^{51}$V($\gamma$,1n) cross section through the refined monochromatic cross section extraction method}

\author{Zi-Rui Hao\orcidC{}}
\affiliation{Shanghai Advanced Research Institute, Chinese Academy of Sciences, Shanghai, 201210, Shanghai, China}

\author{Gong-Tao Fan\orcidD{}}%
\email{fangt@sari.ac.cn}
\affiliation{Shanghai Advanced Research Institute, Chinese Academy of Sciences, Shanghai, 201210, Shanghai, China}
\affiliation{Shanghai Institute of Applied Physics, Chinese Academy of Sciences, 201800, Shanghai, China} \affiliation{University of Chinese Academy of Sciences, 101408, Beijing, China}

 \author{Qian-Kun Sun\orcidE{}}
\affiliation{Shanghai Institute of Applied Physics, Chinese Academy of Sciences, 201800, Shanghai, China}
\author{Hong-Wei Wang\orcidF{}}
 \affiliation{Shanghai Advanced Research Institute, Chinese Academy of Sciences, Shanghai, 201210, Shanghai, China}
 \affiliation{Shanghai Institute of Applied Physics, Chinese Academy of Sciences, 201800, Shanghai, China}
 \affiliation{University of Chinese Academy of Sciences, 101408, Beijing, China}
\author{Hang-Hua Xu\orcidG{}}
 \affiliation{Shanghai Advanced Research Institute, Chinese Academy of Sciences, Shanghai, 201210, Shanghai, China}
  \affiliation{University of Chinese Academy of Sciences, 101408, Beijing, China}
\author{Long-Xiang Liu\orcidH{}}
 \affiliation{Shanghai Advanced Research Institute, Chinese Academy of Sciences, Shanghai, 201210, Shanghai, China}
\author{Yue Zhang\orcidI{}}
 \affiliation{Shanghai Advanced Research Institute, Chinese Academy of Sciences, Shanghai, 201210, Shanghai, China}

\author{Yu-Xuan Yang\orcidK{}}
\affiliation{Shanghai Institute of Applied Physics, Chinese Academy of Sciences, 201800, Shanghai, China}
\author{Kai-Jie Chen\orcidM{}}
\affiliation{Shanghai Institute of Applied Physics, Chinese Academy of Sciences, 201800, Shanghai, China}

 \author{Zhi-Cai Li\orcidQ{}}
 \affiliation{Shanghai Advanced Research Institute, Chinese Academy of Sciences, Shanghai, 201210, Shanghai, China}
\author{Pu Jiao\orcidR{}}
\affiliation{Shanghai Advanced Research Institute, Chinese Academy of Sciences, Shanghai, 201210, Shanghai, China}
\author{Meng-Die Zhou\orcidS{}}
\affiliation{Shanghai Advanced Research Institute, Chinese Academy of Sciences, Shanghai, 201210, Shanghai, China}
\author{Shan Ye\orcidT{}}
\affiliation{Shanghai Advanced Research Institute, Chinese Academy of Sciences, Shanghai, 201210, Shanghai, China}
\author{Zhen-Wei Wang\orcidN{}}
\affiliation{Shanghai Institute of Applied Physics, Chinese Academy of Sciences, 201800, Shanghai, China}
\author{Xiang-Fei Wang\orcidO{}}
\affiliation{Shanghai Institute of Applied Physics, Chinese Academy of Sciences, 201800, Shanghai, China}
\author{Meng-Ke Xu\orcidP{}}
\affiliation{Shanghai Institute of Applied Physics, Chinese Academy of Sciences, 201800, Shanghai, China}

\author{Yu-Long Shen\orcidU{}}
\affiliation{Shanghai Advanced Research Institute, Chinese Academy of Sciences, Shanghai, 201210, Shanghai, China}
\author{Chang Yang\orcidY{}}
\affiliation{Shanghai Advanced Research Institute, Chinese Academy of Sciences, Shanghai, 201210, Shanghai, China}
\author{Jia-Wen Ding\orcidZ{}}
\affiliation{Shanghai Institute of Applied Physics, Chinese Academy of Sciences, 201800, Shanghai, China}

\date{\today}

\begin{abstract}

The Giant Dipole Resonance (GDR) in $^{51}$V has been a long-term conflicting interpretation, with existing photoneutron cross section data suggesting either a single peak or a pronounced splitting, leading to opposite conclusions on nuclear deformation. A new measurement of the $^{51}$V($\gamma$,1n) cross section, performed at the Shanghai Laser Electron Gamma Source (SLEGS) facility, employs a refined monochromatic cross section extraction method. By integrating Polynomial Regression and Support Vector Regression (SVR) for robust interpolation and extrapolation, the new extracted monoenergetic cross sections exhibit a single, broad peak with no evidence of GDR splitting. This result provides new support for a spherical or near-spherical shape of $^{51}$V. Furthermore, we found that deliberately overfitting the data using an SVR model reproduces multi-peak structures similar to those reported in historical datasets, implying that the previously claimed splitting might originated from analysis artifacts rather than physical phenomena. 

\end{abstract}


\maketitle

\section{INTRODUCTION}

The Giant Dipole Resonance (GDR) represents a fundamental collective excitation mode in atomic nuclei, characterized by a broad peak in the photoabsorption cross section arising from the out-of-phase oscillation of protons against neutrons. The structure of the photoabsorption cross section in GDR energy region serves as a direct indicator of nuclear shape: it appears as a single peak in spherical nuclei, while splitting into two distinct components in deformed nuclei due to the energy difference between dipole oscillations along and perpendicular to the deformation axis. This implies that the photoabsorption cross section profile provides crucial experimental insight into nuclear deformation \cite{PhysRevC.71.064328,WOS:000300147700001,JUNGHANS2008200,ZILGES2022103903}. And an accurate and reasonable photoabsorption cross section in GDR energy region is one of the foundations for studying nuclear shape.

The photoabsorption cross section is primarily accessed through photoneutron measurements, since for incident $\gamma$-ray energies above the neutron separation energy $S_{\rm n}$, photon absorption is predominantly followed by neutron emission. The photoneutron cross section not only serves as an essential input for theoretical modeling of nuclear reactions but also provides an approximate measure of the $\gamma$ Strength Function ($\gamma$SF). Under electric dipole ($E$1) excitation, the $\gamma$SF dominates the photoabsorption cross section in the GDR region, making photoneutron data particularly valuable for probing GDR splitting and, consequently, nuclear deformation \cite{WOS:000597810800002}. 

In history, photoneutron cross section in the GDR region has been systematically measured in two laboratories, Livermore and Saclay \cite{dietrich1988atlas}. Unfortunately, a set of 19 nuclei photoneutron cross section data measured from the two laboratories have systematic deviations \cite{WOS:A1975AG97500009,WOS:000339864400001}. Furthermore, significant discrepancies exist in the reported structures of the photoneutron cross sections within the GDR region for certain nuclei; results from some laboratories exhibit a single resonance peak, in contrast to the split peaks reported by the other \cite{VEYSSIERE1974513,PhysRev.128.2345,WOS:A1969E555500025,2019IAEA,WOS:A1970H162800004,WOS:000451576400006}. This is fatal for determining the shape of the nucleus, or whether it has deformed or not. Among these cases, the photoneutron cross section dataset for $^{51}$V in the GDR energy region stands as one of the most emblematic examples of this contradiction.

Three datasets exist for the $^{51}$V($\gamma$,$x$n) reaction \cite{VEYSSIERE1974513,PhysRev.128.2345,WOS:A1969E555500025}. The results from Livermore and Saclay show a similar shape, a single peak, and while a significant global discrepancies. The Saclay data report a higher cross section, while the Livermore data are systematically lower. A comprehensive evaluation by Varlamov et al. suggests that the Livermore measurements suffer from substantial systematic biases, likely due to their use of a 4$\pi$ paraffin-moderated neutron detector. This detector type is prone to misclassifying low-energy neutrons from ($\gamma$,1n1p) reactions as part of the ($\gamma$,2n) channel, leading to an underestimation of the ($\gamma$,1n) cross section and a corresponding overestimation of the ($\gamma$,2n) component. In contrast, the Saclay data, obtained with a gadolinium-loaded liquid scintillator, are considered more reliable, though their ($\gamma$,1n) integrated cross section still falls about 1.8\% below the evaluated value \cite{WOS:000698392600001,WOS:001173151100003,WOS:000339864400001}. The third ($\gamma$,$x$n) dataset for $^{51}$V, measured by B.I. Goryachev \cite{WOS:A1969E555500025} using a bremsstrahlung $\gamma$-ray source, which shows the photoneutron cross section shape in GDR energy region has obvious splitting. Some researchers have interpreted these data as suggesting significant structure in the GDR decay \cite{WOS:000264628400001}, while Varlamov et al. \cite{WOS:001173151100003} consider them unreliable.

Further analysis reveals that there are significant differences in the experimental conditions and data processing methods of the three $^{51}$V datasets. The datasets from Livermore and Saclay were obtained from the 2--4\% quasi-monoenergetic $\gamma$-ray sources. Although both the Livermore and Saclay studies used sources with good energy resolution, neither provided detailed descriptions of their unfolding procedures \cite{VEYSSIERE1974513,PhysRev.128.2345}. While the dataset measured by B.I. Goryachev using the bremsstrahlung, which is a continuous $\gamma$-ray and unfolding procedures must be adopted. Typically, methods such as the Penfold-Leiss approach or the Least Structure Method are applied for this purpose \cite{WOS:001171902400002}. However, due to insufficiently accurate characterization of the bremsstrahlung spectrum near its endpoint energy and increased yield measurement uncertainties above the GDR maximum, the extracted monoenergetic cross sections may have anomalous structures \cite{WOS:001171902400002}. Therefore, when using $\gamma$-ray sources with broad energy spreads for monoenergetic cross section extraction, both the unfolding algorithms and the measured data must be sufficiently precise.

Given the challenges associated with unfolding algorithms discussed above, an unfolding iteration method was developed and has proven effective for monochromatic cross section extraction at 2--7\% resolution, Laser-Compton Back-Scattering (LCBS) facilities like NewSUBARU BL01 beamline \cite{WOS:000412435800003, krzysiek2019photoneutron, WOS:000450548900001}. However, this method proves inadequate for the Shanghai Laser Electron Gamma Source (SLEGS) \cite{AuTbScibull,liu2024energy,xu2022interaction,hao2021collimator}, a Laser Compton Slant Scattering (LCSS) facility, due to its $\gamma$-ray spectrum is significantly different from that of LCBS. The spline interpolation in the unfolding iteration method introduces unphysical oscillations when processing the $\gamma$-ray spectra generated by this slant-scattering approach, potentially creating artificial structures. Owing to the distinct spectral properties arising from LCSS, a reliable extraction at SLEGS requires a method capable of stable interpolation and physically sound extrapolation. To address this,  especially aiming at issue of the $^{51}$V datasets, we introduce a key methodological advancement: a refined unfolding iteration framework incorporating machine learning techniques--specifically Polynomial Regression and Support Vector Regression--to perform stable extrapolation.

In this paper we present a new measurement of the $^{51}$V($\gamma$,1n) cross section, obtained using the refined monochromatic extraction method at SLEGS.

\section{EXPERIMENTAL PROCEDURE}

\begin{figure}
\includegraphics[width=\linewidth]{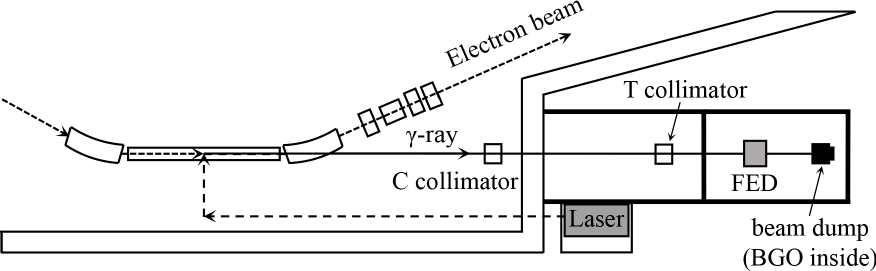}
\caption{\label{experiment_schematic}Schematic diagram of the experimental setup for photoneutron cross section measurements.}
\end{figure}

The measurement was performed at SLEGS beamline (https://cstr.cn/31124.02.SSRF.BL03SSID) of the Shanghai Synchrotron Radiation Facility (SSRF) \cite{000272034800012,WOS:001277786000001} which was operated in top-up mode at 3.5 GeV with a beam current of 220/180 mA. A CO$_2$ laser operating at 5 W, 1 kHz repetition rate, and 50 $\upmu$s pulse width Compton scattering with the SSRF electrons under a slant scattering process to generate turnble $\gamma$-rays in an energy between the neutron separation energy ($S_{\rm{n}}$) and the double neutron separation energy ($S_{\rm{2n}}$). The $\gamma$ beam was aligned using a 5 mm aperture Coarse collimator (C) and a 2 mm aperture Three-hole collimator (T). And then the collimated quasi-monochromatic $\gamma$ beam directed onto a $^{51}$V target, composed of three stacked pieces, with a total thickness of 5.768 mm with a diameter of 10 mm and an isotopic enrichment of 99.95\%.  The experiment set up illustrated in Fig. \ref{experiment_schematic}.

A $^3$He-based Flat Efficiency Detector (FED) was carefully aligned with the target placed at its geometric center to detect the photoneutrons. The FED consists of 26 $^3$He proportional counters embedded in a polyethylene moderator, arranged in three concentric rings at radial distances of 65 mm, 110 mm, and 175 mm from the beam axis. All $^3$He counters have a sensitive length of 500 mm and are filled with 2 atm of $^3$He gas. The inner ring (Ring-1) uses 1-inch diameter counters, while the middle (Ring-2) and outer (Ring-3) rings use 2-inch diameter counters. The inner polyethylene moderator measures 450 mm (W) $\times$ 450 mm (H) $\times$ 550 mm (L). To shield against environmental neutrons, the assembly is wrapped with 2 mm thick cadmium sheets on all six sides, and the entire structure is enclosed by additional polyethylene plates. FED achieves a total neutron detection efficiency of $\sim$42\% \cite{HaoNST2025}. 

\begin{figure}
\centering
	\includegraphics[width=\linewidth]{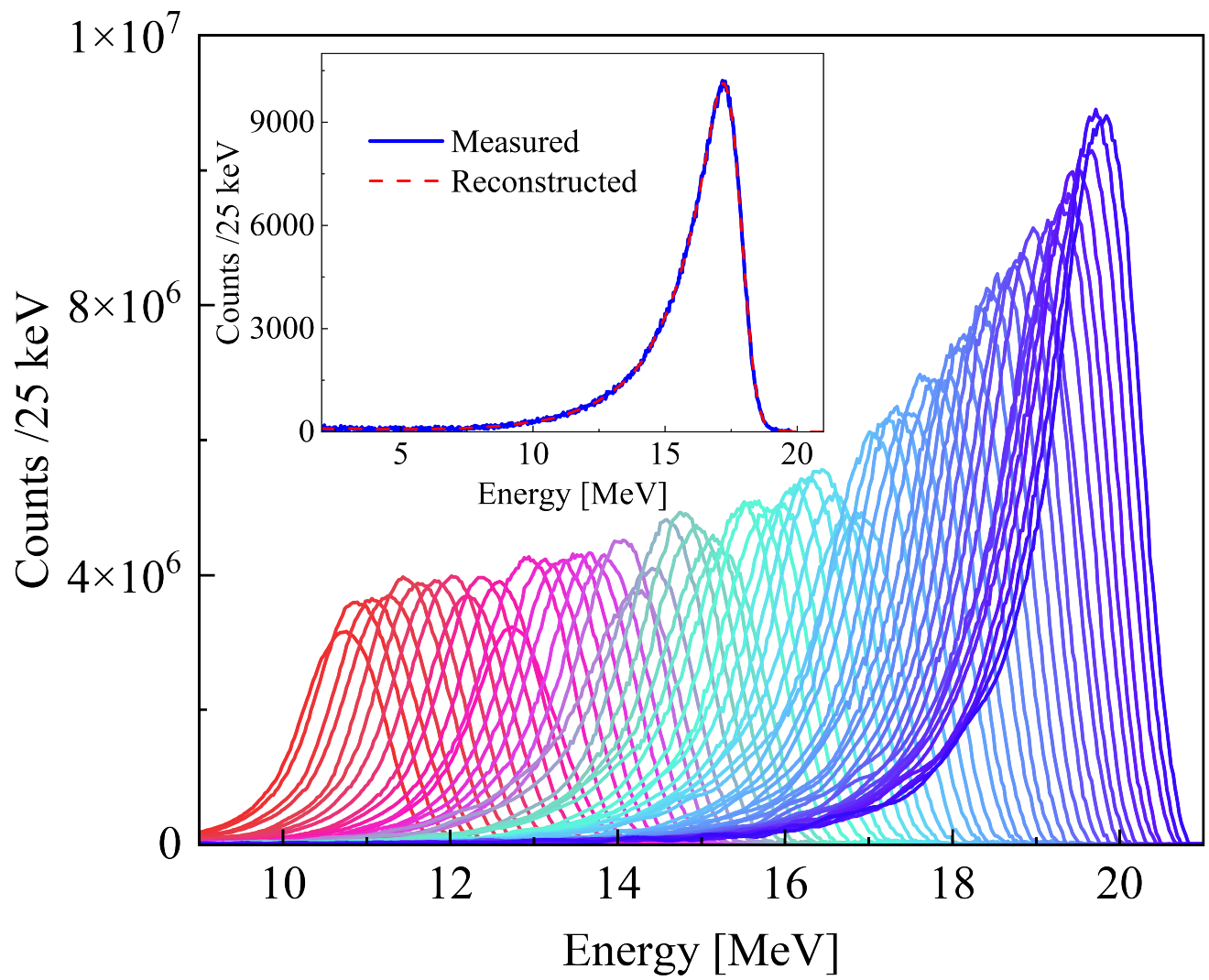}
\caption{\label{AllGammaSpectra} The incident $\gamma$-ray spectra used in the measurement, normalized to one hour. The consistency between the measured and reconstructed spectra for a representative case is exemplified in the inset.
 }
\end{figure}

The $\gamma$-ray beam was measured with a $\Phi$76 mm $\times$ 200 mm BGO detector after passing through the $^{51}$V target. The detector, which has an efficiency of approximately 100\%, was carefully calibrated via proton capture reactions at the tandem accelerator \cite{liu2024energy}. To match the counting rate conditions of the calibration experiment, a copper attenuator was used to reduce the online counting rate to approximately 1 kHz. The incident $\gamma$-ray spectra were reconstructed using the Direct Unfolding Method \cite{liu2024energy}. The normalized $\gamma$-ray spectra are shown in Fig. \ref{AllGammaSpectra}.

\section{The refined Unfolding Framework}

\subsection{The monochromatic cross section extraction method}\label{section3.A}

The folded photoneutron cross section can be performed as
\begin{equation}
\int_{S_{\rm n}}^{E_{\rm max}} n_{\gamma}(E_\gamma)\sigma_{\gamma \rm n}(E_\gamma)\rm{d} \it{E_\gamma} = \frac{N_{\rm n}}{N_{\rm t}t_\gamma\xi\epsilon_{\rm n}g}
\label{crosssectiondefination},
\end{equation}

\noindent
where, the $n_{\gamma}(E_\gamma)$ is the normalized energy distribution of the incident $\gamma$-rays. The $\sigma_{\gamma \rm n}(E_\gamma)$ is the monoenergetic (unfolded) cross section which will be determined. $N_{\rm n}$ is the number of detected photoneutrons. $N_\gamma$ is the number of $\gamma$ rays incident on the target. For detailed procedures on the extraction of $N_{\rm n}$ and $N_\gamma$, please refer to the reference \cite{HAO2025171026}. $\epsilon_{\rm n}$ represents the average detector efficiency, determined by the counting ratio between the outer ring and the inner ring, namely the Ring-Ratio technique. $N_{\rm t}$ is the number of target nuclear per unit area. $ \xi=(1-e^{-\mu t})/(\mu t) $ is a correction factor for target self-attenuation. $\mu$ is the attenuation coefficient for $\gamma$ rays. $t$ is the target thickness. $g$ is the fraction of the $\gamma$ flux above the $S_{\rm n}$,

\begin{equation}
g=\frac{\int_{S_{\rm n}}^{E_{\rm max}}n_\gamma(E_\gamma)\rm{d} \it{E_\gamma}}{\int_{0}^{E_{\rm max}}n_\gamma(E_\gamma)\rm{d} \it{E_\gamma}}
\label{crosssectionfactorg}.
\end{equation}

The right side of Eq. \ref{crosssectiondefination} represents the folded cross section, which is determined directly from the experimental data. It can be known as the convolution of the monochromatic cross sections with the normalized incident $\gamma$-ray spectrum, representing a weighted average over the energy distribution.

The folded cross sections were measured at a series of separate energies. Consequently, these quantities are connected to the monochromatic cross sections through the following linear relation:
\begin{equation}
\sigma_{\rm{f}}= \textbf{D}\sigma
\label{monocrosssectionandfoldedCS}.
\end{equation}

\noindent

$\sigma_{\rm{f}}$ represents a set of folded cross sections, with each element corresponding to the folded cross section measured at specific discrete beam energies ($E_\gamma$). Here, $\sigma$ denotes the array of unfolded cross sections. Each row of the matrix \textbf{D} represents a normalized incident $\gamma$ energy spectrum ranging from $S_{\rm{n}}$ to $E_{\text{max}}$. Equation \ref{monocrosssectionandfoldedCSexpended} is the expansion form of Eq. \ref{monocrosssectionandfoldedCS}. The number of rows ($N$) in \textbf{D} corresponds to the number of discrete beam energies, while the number of columns ($M$) corresponds to the number of bins in the incident $\gamma$ spectrum from $S_{\rm{n}}$ to $E_{\text{max}}$.

\begin{equation}
\begin{pmatrix} 
\sigma_1\\
\sigma_2\\
\vdots\\
\sigma_N\
\end{pmatrix}_{\rm{f}}=
\begin{pmatrix} 
D_{11} & D_{12} & \dots &D_{1M} \\
D_{21} & D_{22} & \dots &D_{2M} \\
\vdots & \vdots & \ddots & \vdots \\
D_{N1} & D_{N2} & \dots & D_{NM} \
\end{pmatrix}
\begin{pmatrix} 
\sigma_1\\
\sigma_2\\
\vdots\\
\vdots\\
\sigma_M\
\end{pmatrix}
\label{monocrosssectionandfoldedCSexpended}.
\end{equation}

An unfolding iteration technique is employed to extract the monochromatic cross section $\sigma$. Here is a concise overview:

(1) For the initial iteration, $\sigma^0$ is initialized as a constant array. Consequently, $\sigma_{\rm{f}}^0$ can be calculated using Equation \ref{monocrosssectionandfoldedCS}.

(2) For the subsequent iteration, $\sigma^1$ is derived by adding the deviation between $\sigma_{\rm{exp}}$ and $\sigma_{\rm{f}}^0$ to $\sigma^0$. Given that the dimension of $\sigma^{0}$ ($M=390$) is considerably larger than that of $\sigma_{\rm{exp}}$ and $\sigma_{\rm{f}}^0$ ($N=59$), an interpolation procedure is individually applied to $\sigma_{\rm{exp}}$ and $\sigma_{\rm{f}}^0$ to expand their dimensions to $M$.

\begin{equation}
\sigma^{1}= \sigma^{0}+(\sigma_{\rm{exp}}-\sigma_{\rm{f}}^0)
\label{interpolation1}.
\end{equation}
\noindent

(3) The $i$-th iteration can be concluded as,

\begin{eqnarray}
\sigma_{\rm{f}}^i & = & \textbf{D}\sigma^i ,
\\
\sigma^{i+1} & = & \sigma^{i}+(\sigma_{\rm{exp}}-\sigma_{\rm{f}}^i).
\label{interpolationi}
\end{eqnarray}

The Mean Squared Error (MSE) between $\sigma_{\rm{exp}}$ and $\sigma_{\rm{f}}^{i+1}$ is calculated in each iteration until convergence is achieved. The resulting $\sigma^{i}$ is then taken as the unfolded cross section array, with the difference between $\sigma^{i+1}_{\rm f}$ and $\sigma_{\rm{exp}}$ being smaller than the statistical uncertainty.

\subsection{The interpolation procedure in unfolding iteration method}

In step (2) of the unfolding iteration procedure, the dimension of $\sigma_{\rm{exp}}$ and $\sigma_{\rm{f}}^i$ is enlarged to match that of $\sigma^{i}$ via interpolation. As implemented at NewSUBARU BL01 beamline, this is done with spline interpolation \cite{WOS:000450548900001}. Equation \ref{monocrosssectionandfoldedCSexpended} shows that the incident $\gamma$-ray energy spans from $S_{\rm n}$ to $E_{\rm {max}}$. However, the lowest measured energy point, $E_{(\sigma_1)_{\rm f}}$ (corresponding to $(\sigma_1)_{\rm f}$ in Eq. \ref{monocrosssectionandfoldedCSexpended}), is actually greater than $S_{\rm n}$. Similarly, the highest measured energy point, $E_{(\sigma_N)_{\rm f}}$ (corresponding to $(\sigma_N)_{\rm f}$ in Eq. \ref{monocrosssectionandfoldedCSexpended}), is less than $E_{\rm {max}}$. Therefore, during the dimension extension step, reasonable extrapolation is required for the intervals  $S_{\rm n}$--$E_{(\sigma_1)_{\rm f}}$ and $E_{(\sigma_N)_{\rm f}}$--$S_{\rm n}$, which lie outside the directly measured range, to ensure the physical validity of the extracted monoenergetic cross sections. Unfortunately, spline interpolation is generally weak in extrapolation capability and often introduces polynomial oscillation.

At the NewSUBARU BL01 beamline, the photoneutron cross section measurements benefit from excellent energy resolution and a theoretical energy (in back-scattering mode) near the maximum \cite{HORIKAWA2010209}. These advantages make spline interpolation sufficient for extending the dimensions of $\sigma_{\rm{exp}}$ and $\sigma_{\rm{f}}^i$. In contrast, for SLEGS, its moderate $\gamma$-ray energy resolution and the use of slant-scattering mode for $\gamma$-ray production result in significant intervals between $S_{\rm n}$ and $E_{(\sigma_1)_{\rm f}}$, and between $E_{(\sigma_N)_{\rm f}}$ and $E_{\rm {max}}$. Extrapolation over these regions is necessary, rendering the conventional spline interpolation method unsuitable for SLEGS data analysis.

To achieve extrapolation, SLEGS employs machine learning methods such as Polynomial Regression and Support Vector Regression (SVR). The following will present the results of extracting the unfolded cross section of $^{51}$V($\gamma$,n)$^{50}$V using the unfolding iteration method integrated with these machine learning techniques.

Polynomial Regression offers an intuitive and computationally efficient means of nonlinear fitting, while SVR provides robust handling of complex nonlinear relationships through its kernel trick. Although numerous more complex machine learning models exist that could perform the "dimension extension" (extrapolation) task for spectral data, the specific problem addressed in this study features limited data scale and relational complexity. Therefore, the two selected methods are fully sufficient to deliver accurate and stable extrapolation results while guaranteeing high computational efficiency.

The key tunable parameter for the polynomial regression in this work is the polynomial degree, which dictates the model's flexibility. An insufficient degree leads to underfitting, failing to capture the underlying trend, while an excessively high degree causes overfitting, producing unphysical oscillations upon extrapolation. Although regularization techniques such as Ridge or Lasso regression can, in principle, mitigate overfitting by penalizing large coefficients in the loss function, they were found to be unsuitable for the present dataset. Empirical evaluation showed that regularization overly constrained the model, preventing it from effectively capturing the genuine cross section trend. Consequently, the standard Linear Regression algorithm was employed.

\begin{figure}
\centering
	\includegraphics[width=\linewidth]{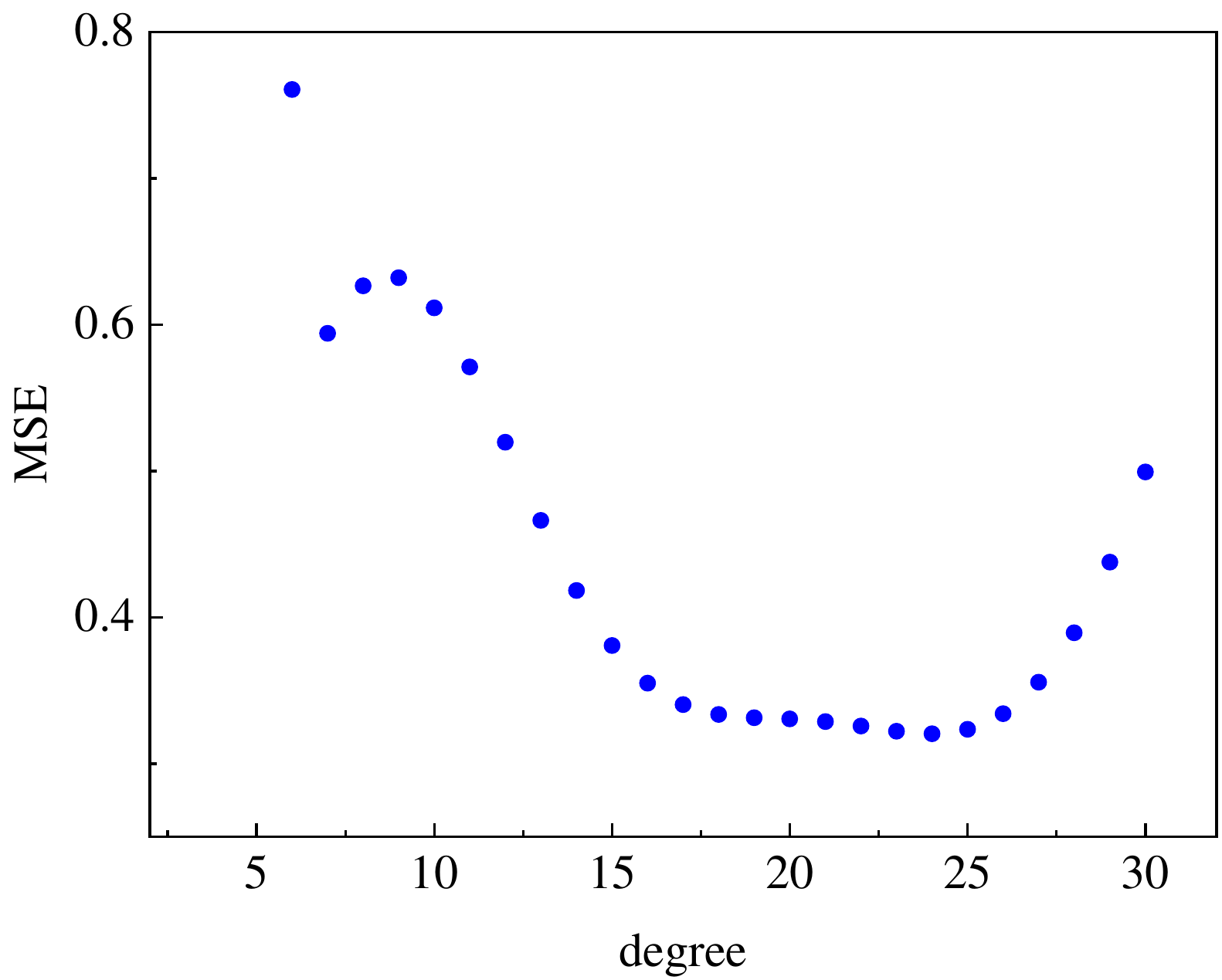}
\caption{\label{convergence_MSE_degree} The converged MSE between $\sigma_{\rm{exp}}$ and $\sigma_{\rm{f}}^{i+1}$ as a function of the polynomial degree.  }
\end{figure}

\begin{figure}
\centering
	\includegraphics[width=\linewidth]{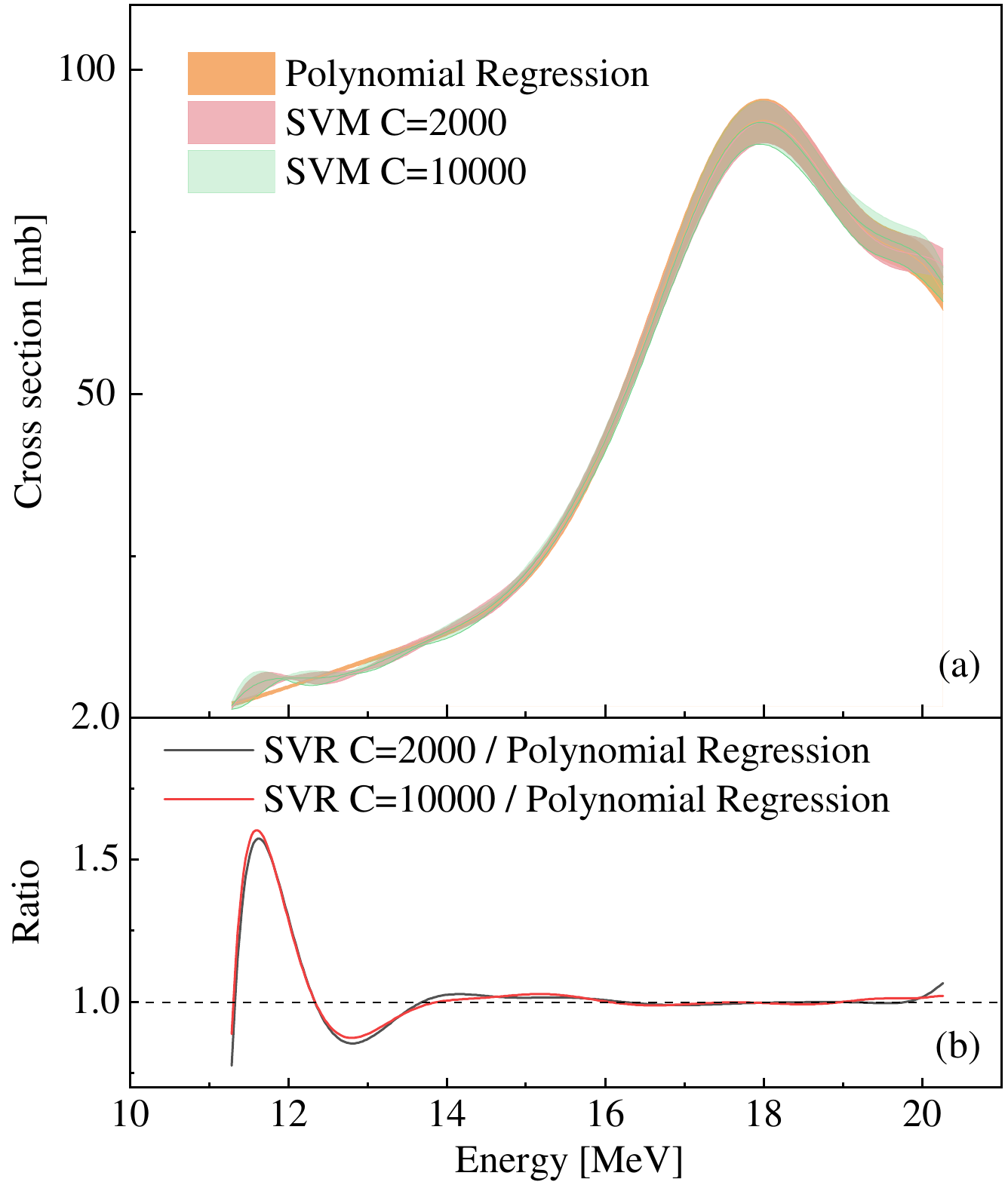}
\caption{\label{CrossSectionbyvariousmodels_ratios} Unfolded cross section for $^{51}$V($\gamma$,n)$^{50}$V (a) and its ratios (b) obtained from Polynomial Regression and SVR models with two different regularization strengths $\it{C}$. }
\end{figure}

Therefore, for the polynomial regression, only the degree required optimization. Figure \ref{convergence_MSE_degree} shows the relationship between the final converged MSE and the polynomial degree. The MSE reached its global minimum at a degree of 24. When the degree was significantly lower than 24 (e.g., in the range of 5 to 15 as shown in the figure), the polynomial regression failed to adequately capture the trend of the cross sections, suggesting underfitting. Conversely, when the degree greatly exceeded 24, the results exhibited multi-peak artifacts, indicating overfitting. The unfolded cross sections using this optimal parameter are therefore expected to be a closer approximation to the true cross sections. The resultant unfolded cross sections are shown in Fig. \ref{CrossSectionbyvariousmodels_ratios}(a) by the orange band with the band represented as the total uncertainties.

As for the SVR, the hyperparameters include $\it{kernel}$, $\it{\Gamma}$ and the penalty parameter $\it{C}$. The choice of $\it{kernel}$ determines SVR's capability to map data into a higher-dimensional feature space. The Radial Basis Function (RBF) kernel is most common for handling highly nonlinear relationships, whose effectiveness is governed by the $\it{\Gamma}$ parameter. The $\it{\Gamma}$ defines the influence range of a single training sample. A low $\it{\Gamma}$ value implies a larger influence radius, leading to a smoother and simpler decision boundary, which helps prevent overfitting. A high $\it{\Gamma}$ value makes the model closely fit individual data points, resulting in a complex, wiggly boundary that is risky for extrapolation. The $\it{C}$ controls the model's tolerance for errors. A small $\it{C}$ emphasizes a "maximum margin" boundary, tolerating more errors and yielding a simpler model. A large $\it{C}$ forces the model to fit the data more strictly, potentially reducing its generalization capability.

\begin{figure}
\centering
	\includegraphics[width=\linewidth]{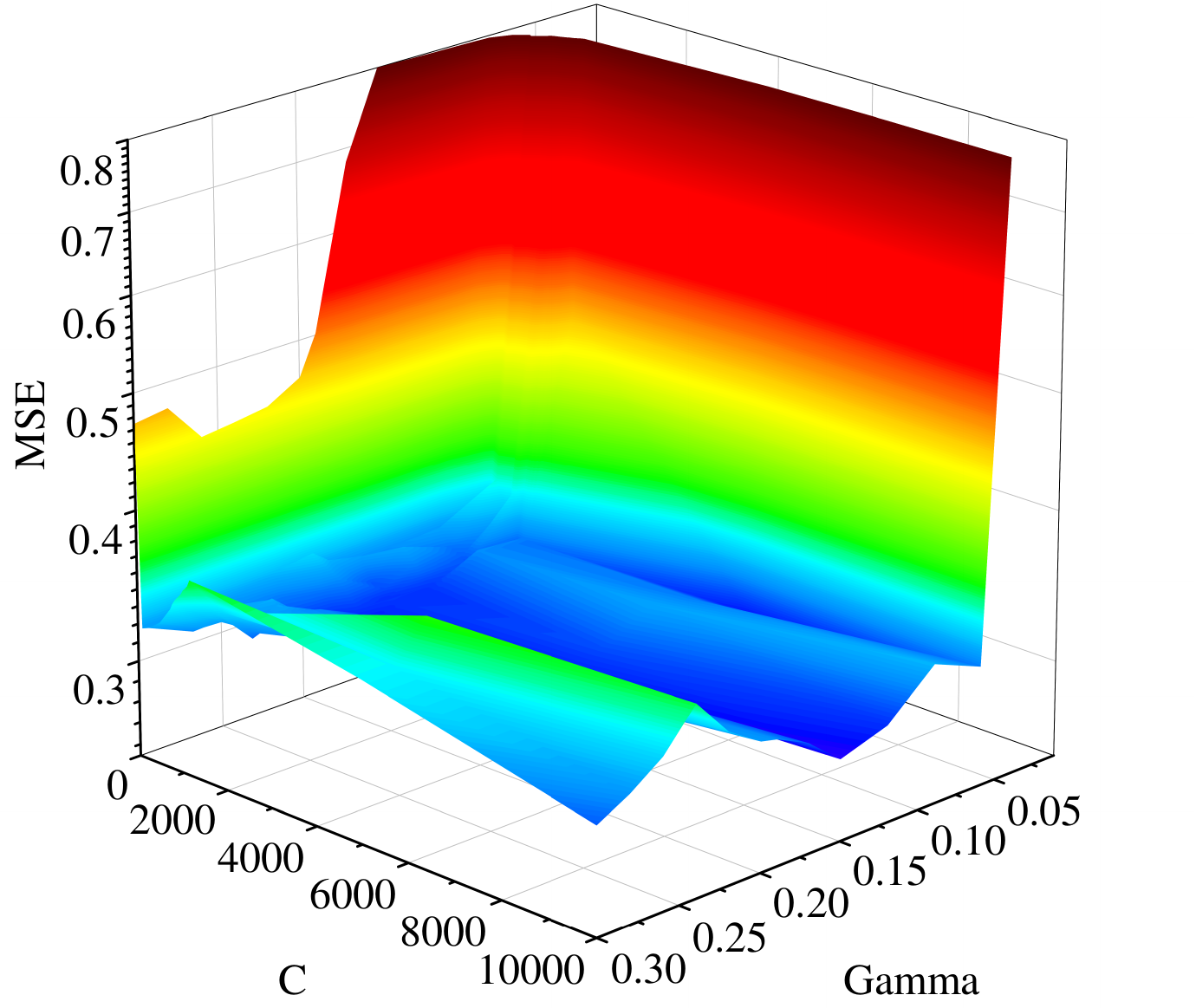}
\caption{\label{convergence_MSE_C_gamma}The final converged MSE values of the unfolding iteration for different combinations of the SVR hyperparameters $\it{C}$ and $\it{\Gamma}$ (RBF kernel). A region of optimal performance (MSE $\approx$ 0.3) is found for $\it{\Gamma}$ = 0.15 and $\it{C}$ between 2000 and 10000. }
\end{figure}

The convergence of the unfolding iteration is monitored by the MSE between $\sigma_{\rm{exp}}$ and $\sigma_{\rm{f}}^{i+1}$, as described previously. This MSE value converges as the iteration proceeds, reaching distinct final values for different combinations of the hyperparameters $\Gamma$ and $C$. Figure \ref{convergence_MSE_C_gamma} illustrates the relationship between the MSE convergence values and the chosen values of $\it{\Gamma}$ and $\it{C}$. It can be observed that when either $\it{\Gamma}$ or $\it{C}$ falls below a certain threshold, the MSE increases rapidly, indicating that the model fails to describe the data adequately. Conversely, the MSE also increases as $\it{\Gamma}$ and $\it{C}$ become too large, which causes the model to overfit the data and produce spurious peaks. The lowest MSE range, approximately 0.3, is achieved when $\it{\Gamma}$=0.15 and $\it{C}$ is within the range of 2000 to 10000. Consequently, we performed the unfolded cross section extraction using the parameter sets ($\it{\Gamma}$, $\it{C}$) = (0.15, 2000) and (0.15, 10000), which yielded the optimal convergence. The results are represented by the red and green bands, respectively, in Fig. \ref{CrossSectionbyvariousmodels_ratios}(a), where the bandwidth denotes the total uncertainty.

A comparison of the unfolded cross sections obtained from the Polynomial Regression and SVR methods is presented in Fig. \ref{CrossSectionbyvariousmodels_ratios}(a). While both models exhibit strong agreement in the overall trend, the SVR result displays significant oscillations below 14 MeV. The ratio of the SVR-derived cross section to the Polynomial Regression result, shown in Fig. \ref{CrossSectionbyvariousmodels_ratios}(b), clarifies this behavior, revealing pronounced fluctuations in the SVR solution in the low-energy region. Physically, an enhancement near the reaction threshold could suggest contributions from a pygmy resonance. However, the oscillatory pattern of the ratio, which eventually converges to unity, is more indicative of an artifact introduced by the SVR model during the unfolding process rather than a genuine resonance structure. Therefore, we conclude that the unfolded cross section derived from the Polynomial Regression-based unfolding iteration is more physically reasonable and reliable.

\section{Results and Discussion }

\subsection{The unfolded cross section}

\begin{figure}
\centering
	\includegraphics[width=\linewidth]{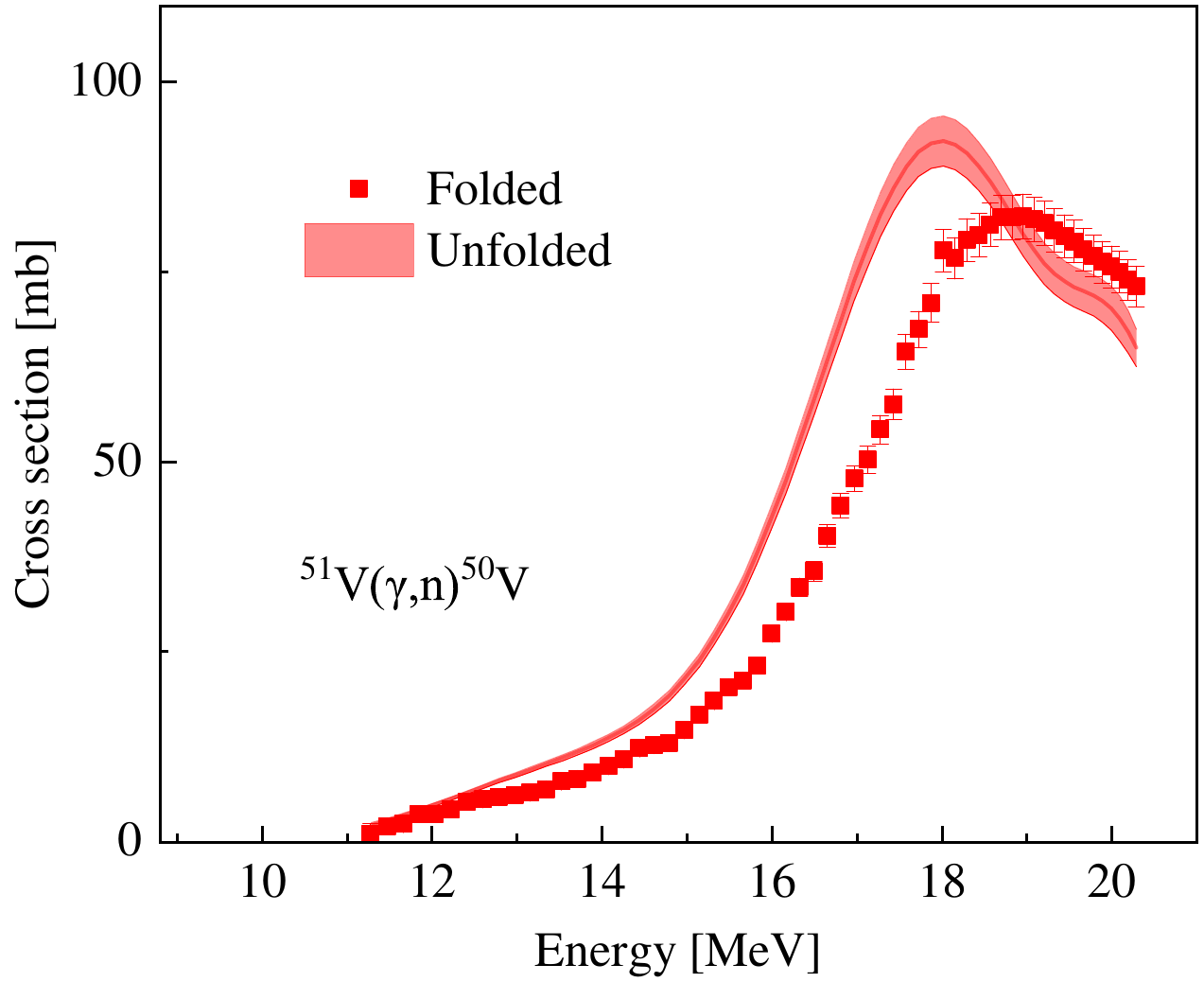}
\caption{\label{quasimonochromaticcrosssectionsandmonochromaticcrosssections_51V}The folded and unfolded cross sections of $^{51}$V($\gamma$,n)$^{50}$V.}
\end{figure}

The comparison between the folded and the unfolded cross sections are shown in Fig. \ref{quasimonochromaticcrosssectionsandmonochromaticcrosssections_51V}. The error bars in these figures represent the total uncertainties, which include the statistical uncertainties, the systematic uncertainties, the methodological uncertainties. The statistical uncertainties arise solely from neutron counting, as the abundant $\gamma$-ray counts make their associated uncertainty negligible. Methodological uncertainties consist of two components: first, the manual setting of parameters in $N_{\rm n}$ extraction introduces fluctuations of less than 2\% in the extracted neutron yield; second, the response matrix used to reconstruct the incident $\gamma$-ray spectrum carries an uncertainty of $\sim$1\%, which propagates to a $\sim$1\% uncertainty in the determined $\gamma$-ray flux. Systematic uncertainties comprise three contributions: (1) The total uncertainty in the FED efficiency is 3.02 \%. (2) The uncertainty in the reconstructed incident energy spectrum introduced by the external copper attenuator and the target is 0.90 \%. (3) The uncertainty attributable to target thickness is estimated to be $\textless $ 0.10\%. Uncertainties of the folded cross section are propagated through standard error-propagation formulas. For the unfolded cross section, the upper and lower bounds of the folded cross section uncertainty are each fed into the unfolding iteration procedure to determine the corresponding unfolded uncertainty. The cross section near the $S_{\rm n}$ region is small, so statistical uncertainty dominates and drives the total uncertainty higher; away from $S_{\rm n}$ the typical total uncertainty is 2--3\%. The monochromatic cross sections and the corresponding uncertainties are provided in the Table \ref{tab:V51}.

\begin{table}
\caption{\label{tab:V51}The monochromatic cross sections of $^{51}$V($\gamma$,n)$^{50}$V.  }
\begin{ruledtabular}
\begin{tabular}{cccccc}
\makecell[c]{$E_\gamma$ \\ $[\text{MeV}]$}& \makecell[c]{$\sigma_{\gamma \rm n}$\\ $[\text{mb}]$}& \makecell[c]{$\sigma_{\gamma \rm n}^{\rm {Stat}}$ \\ $[\text{mb}]$} & \makecell[c]{$\sigma_{\gamma \rm n}^{\rm {Meth}}$ \\ $[\text{mb}]$ } & \makecell[c]{$\sigma_{\gamma \rm n}^{\rm {Syst}}$ \\ $[\text{mb}]$ }& \makecell[c]{$\sigma_{\gamma \rm n}^{\rm {Total}}$ \\ $[\text{mb}]$}\\
\hline

11.28&2.16&0.36&0.04&0.07&0.37\\
11.47&2.81&0.31&0.05&0.09&0.33\\
11.66&3.49&0.25&0.06&0.11&0.28\\
11.85&4.21&0.20&0.07&0.13&0.25\\
12.03&4.96&0.14&0.08&0.16&0.23\\
12.22&5.72&0.09&0.09&0.18&0.22\\
12.41&6.49&0.05&0.11&0.20&0.23\\
12.60&7.26&0.01&0.12&0.23&0.26\\
12.78&8.03&0.03&0.13&0.25&0.28\\
12.97&8.79&0.05&0.14&0.28&0.32\\
13.16&9.54&0.06&0.15&0.30&0.34\\
13.34&10.29&0.07&0.16&0.32&0.36\\
13.53&11.06&0.06&0.18&0.35&0.40\\
13.71&11.86&0.05&0.19&0.37&0.42\\
13.89&12.72&0.02&0.20&0.40&0.45\\
14.07&13.67&0.01&0.22&0.43&0.48\\
14.25&14.75&0.05&0.23&0.46&0.52\\
14.43&16.02&0.09&0.26&0.50&0.57\\
14.61&17.51&0.13&0.28&0.55&0.63\\
14.79&19.28&0.17&0.31&0.61&0.71\\
14.96&21.39&0.20&0.34&0.67&0.78\\
15.14&23.87&0.23&0.38&0.75&0.87\\
15.31&26.76&0.24&0.43&0.84&0.97\\
15.48&30.08&0.25&0.48&0.95&1.09\\
15.66&33.84&0.24&0.54&1.07&1.22\\
15.82&38.03&0.23&0.61&1.20&1.37\\
15.99&42.63&0.20&0.68&1.34&1.52\\
16.16&47.55&0.17&0.76&1.50&1.69\\
16.32&52.74&0.14&0.85&1.66&1.87\\
16.49&58.09&0.11&0.93&1.83&2.06\\
16.65&63.47&0.08&1.02&2.00&2.25\\
16.81&68.75&0.07&1.11&2.17&2.44\\
16.96&73.79&0.06&1.20&2.33&2.62\\
17.12&78.45&0.07&1.28&2.47&2.78\\
17.27&82.58&0.10&1.35&2.60&2.93\\
17.43&86.07&0.14&1.41&2.71&3.06\\
17.58&88.82&0.19&1.46&2.80&3.16\\
17.72&90.77&0.24&1.50&2.86&3.24\\
17.87&91.88&0.30&1.52&2.90&3.29\\
18.01&92.18&0.34&1.52&2.91&3.30\\
18.15&91.71&0.36&1.52&2.89&3.29\\
18.29&90.56&0.37&1.50&2.85&3.24\\
18.43&88.87&0.34&1.47&2.80&3.18\\
18.57&86.79&0.28&1.43&2.74&3.10\\
18.70&84.49&0.20&1.39&2.66&3.01\\
18.83&82.13&0.09&1.36&2.59&2.93\\
18.95&79.87&0.02&1.32&2.52&2.84\\
19.08&77.83&0.11&1.30&2.45&2.78\\
19.20&76.11&0.18&1.28&2.40&2.73\\
19.32&74.75&0.19&1.28&2.36&2.69\\
19.44&73.74&0.13&1.29&2.33&2.67\\
19.55&73.00&0.02&1.31&2.30&2.65\\
19.67&72.42&0.27&1.33&2.29&2.66\\
19.78&71.87&0.58&1.36&2.27&2.71\\
19.88&71.17&0.94&1.38&2.25&2.80\\
19.99&70.20&1.26&1.39&2.22&2.91\\
20.09&68.84&1.44&1.36&2.17&2.94\\
20.19&67.07&1.32&1.30&2.12&2.82\\
20.28&64.99&0.72&1.18&2.05&2.47\\

\end{tabular}
\end{ruledtabular}
\end{table}

\begin{figure}
\centering
	\includegraphics[width=\linewidth]{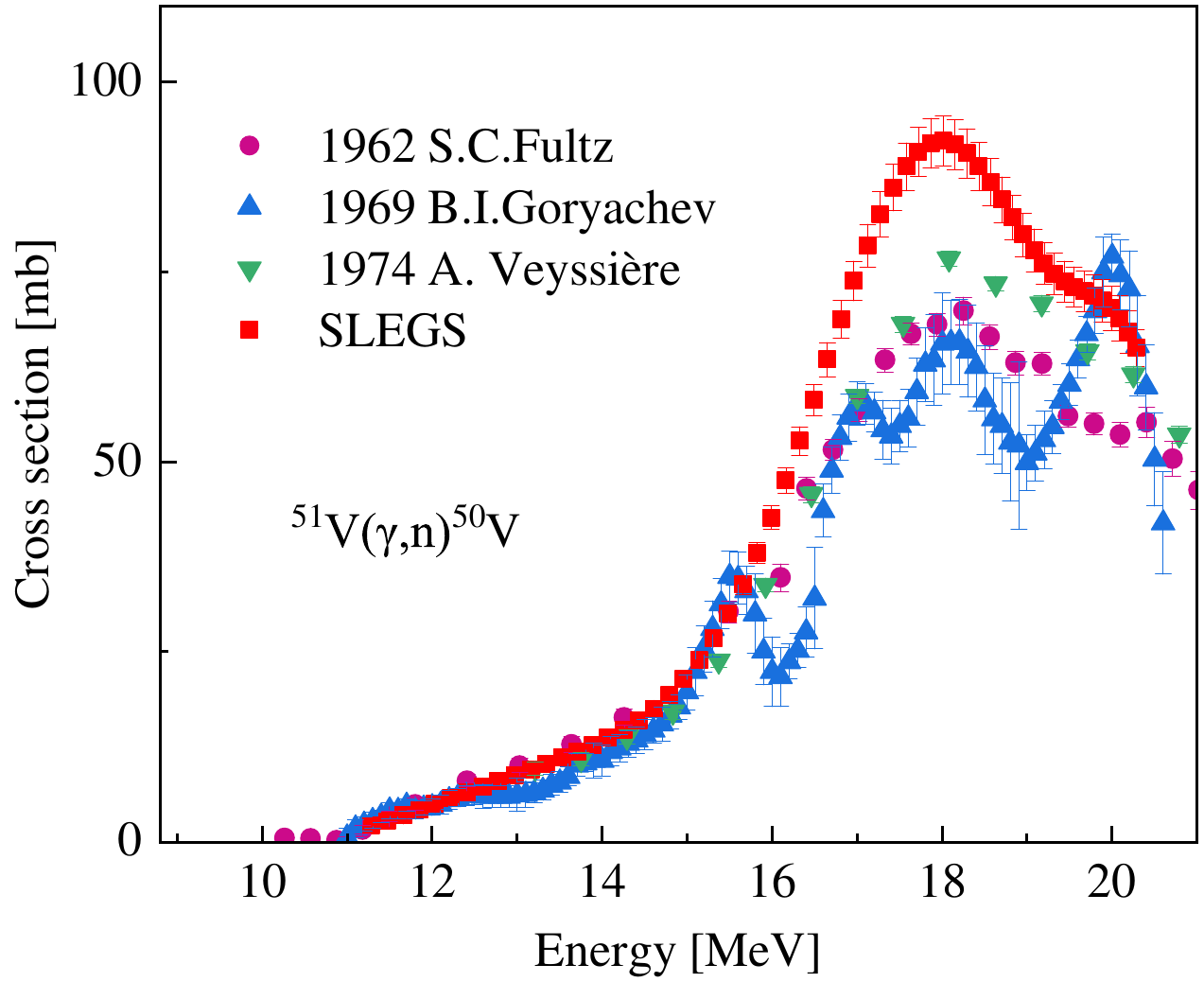}
\caption{\label{V51unfoldedCrossSection} The cross sections for $^{51}$V($\gamma$,n)$^{50}$V in comparison with data from other laboratories.}
\end{figure}

As shown in Fig. \ref{V51unfoldedCrossSection}, the present data exhibit a consistent single-peak shape with the positron in-flight annihilation results from Saclay \cite{VEYSSIERE1974513} and Livermore \cite{PhysRev.128.2345}. In terms of magnitude, our results agree with the Saclay data below 16 MeV but are higher above this energy, and are generally higher than the Livermore data. A slight low-energy shift of the GDR peak is also noted in our data due to the unfolding procedure. A more dramatic discrepancy is apparent with the bremsstrahlung data \cite{WOS:A1969E555500025}, which agree with ours below 12.5 MeV but are lower in the 12.5--15 MeV region and, critically, display a split-peak structure above 15 MeV---a feature not seen in our work or the other datasets.

\subsection{The possible origin of the reported splitting}

The apparent GDR splitting reported in B.I. Goryachev et al. presents a persistent puzzle. We note that such multi-peak structures show a resemblance to spurious oscillations known to arise from overfitting in unfolding procedures. This observation led us to formally test the hypothesis that the reported splitting could be an artifact of the data analysis method rather than a true physical phenomenon.

\begin{figure}
\centering
	\includegraphics[width=\linewidth]{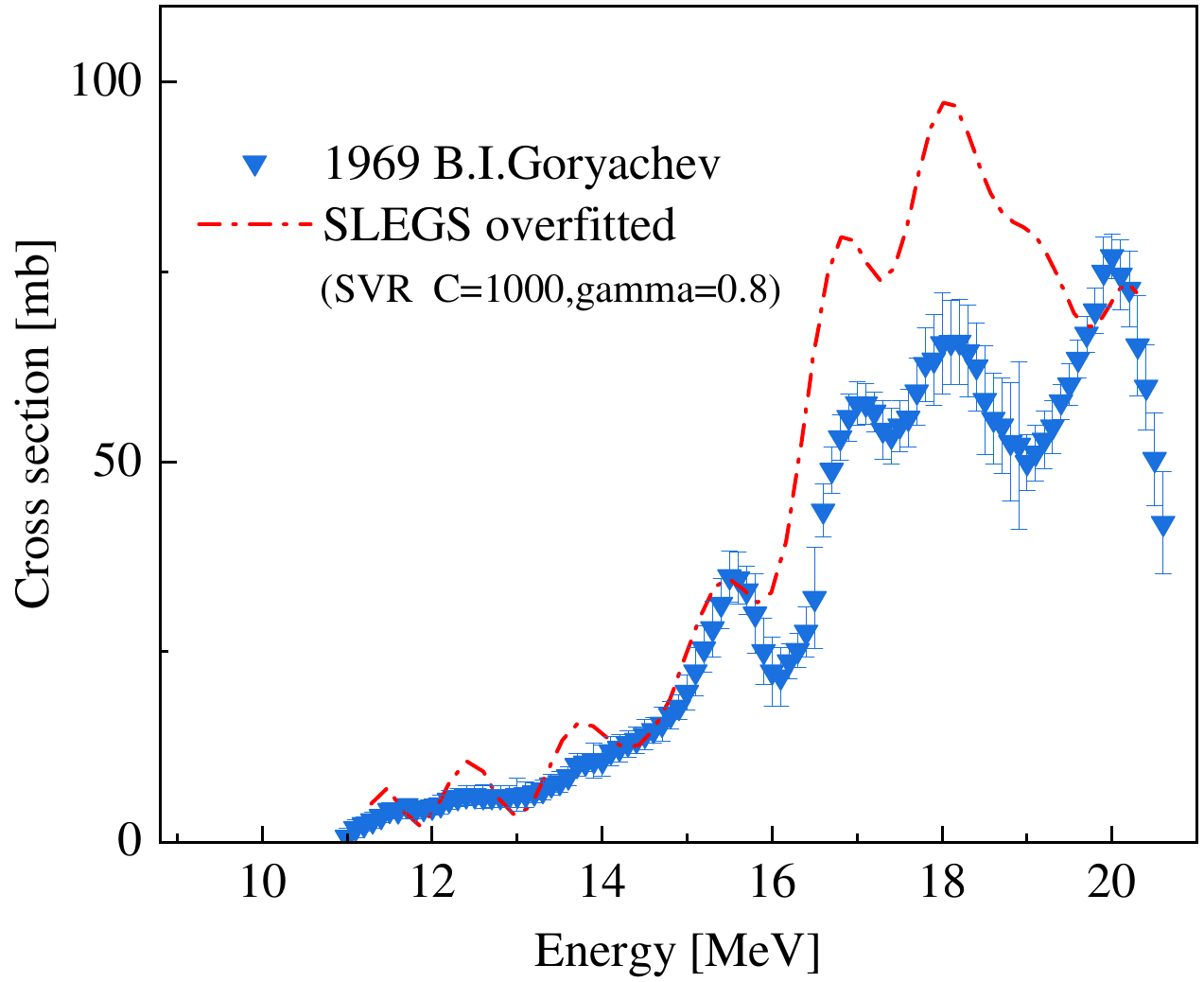}
\caption{\label{ComparewithGoryachev} A comparison of the unfolded cross section extracted using SVR (with parameters $\it{C}$ = 1000, $\it{\Gamma}$=0.8) against the results from B.I. Goryachev (1969). The two methods show agreement in the peak positions of several higher-lying resonances.
}
\end{figure}

To investigate this, we deliberately configured a SVR model with hyperparameters (specifically, a high $\it{\Gamma}$=0.8 and the penalty parameter $\it{C}$=1000) that promote overfitting. When applied within our unfolding iteration, this intentionally misguided SVR model produced a solution riddled with unphysical, multi-peak structures, as shown in Fig. \ref{ComparewithGoryachev}.

Strikingly, the artificially generated structure from this overfitted model (Fig. \ref{ComparewithGoryachev}) exhibits a remarkable similarity in the energy positions of the peaks to those reported in the historical data from B.I. Goryachev et al. This reveals that the specific multi-peak structure in the $^{51}$V($\gamma$,n) cross section may originate from the unfolding algorithm used in the data analysis. While a definitive assessment of the original analysis is impossible, our deliberate numerical test provides a plausible scenario for its origin. It establishes that unfolding overfitting is a possible, and likely, explanation for the reported splitting. Therefore, when combined with the clean, single-peak result obtained via the robust polynomial regression method, the historical multi-peak structures are most likely not representative of physical features.

\section{Conclusion}

In summary, we have performed a new measurement of the $^{51}$V($\gamma$,1n) cross section using a laser Compton slant-scattering $\gamma$-ray source. By integrating machine-learning techniques, specifically polynomial regression, into a refined unfolding method, we have reliably extracted the monoenergetic photoneutron cross section for $^{51}$V. Our results unambiguously reveal a single broad peak in the GDR energy region, same as the results from positron in-flight annihilation ones measured at Saclay and at Livermore, providing clear new evidence against the previously suggested splitting. This finding resolves the long-standing ambiguity regarding the GDR structure in $^{51}$V and strongly supports a spherical or near-spherical ground state shape for this nucleus. Furthermore, we have demonstrated that the multi-peak structures reported in historical data can be reproduced through the deliberate overfitting of our data using an SVR algorithm. This provides a methodology-driven explanation for the origin of the historical controversy, attributing the apparent splitting to an analysis procedure rather than a physical phenomenon. 

The refined unfolding method not only enables reliable unfolded cross section extraction at LCSS $\gamma$-ray facilities with moderate energy resolution but is also applicable to experiments conducted with fine-resolution $\gamma$ beams. Its capability to identify global trends and reject anomalous fluctuations establishes it as a more general and robust approach. Consequently, it sets a robust standard for determining monoenergetic photoneutron cross sections, which is of fundamental importance for ascertaining nuclear deformation and structure.

\section*{ACKNOWLEDGEMENT}

This work was supported by the National key R\&D program (No.2023YFA1606901, No.2022YFA1602400), the National Natural Science Foundation of China (No.12388102, No.12275338, No.U2441221), Shanghai Oriental Talents-Leading Project (No.LJ2024080).

\end{document}